\documentclass[aps,prl,groupedaddress,showpacs]{revtex4-1}

\usepackage{graphicx}% Include figure files
\usepackage{dcolumn}% Align table columns on decimal point
\usepackage{bm}% bold math

\begin{document}

%\preprint{}

\title{Structural Origin of Light Emission in Germanium Quantum Dots}


\author{A. V. Sapelkin}
\email[]{a.sapelkin@qmul.ac.uk}
\author{D. Bolmatov}
\author{A. Karatutlu}
\author{W. Little}
\author{K. Trachenko}
%\thanks{}
\affiliation{Center for Condensed Matter and Materials Physics, School of Physics and Astronomy, Queen Mary, University of London, Mile End Road, London E1 4NS, UK}
\author{G. Cibin}
\author{R. Taylor}
\author{F. Mosselmans}
\author{A. J. Dent}
\affiliation{Diamond Light Source Ltd, Didcot OX11 0DE, Oxon, UK}

\author{G. Mountjoy}
\affiliation{School of Physical Sciences, University of Kent, Canterbury, CT2 7NZ, UK}


\date{\today}

\begin{abstract}
The origin of visible light emission from nanostructures has been a subject of an intense debate since the early work by L. E.  Brus and A. P. Alivisatos in 1980s. The intense research that followed has paved the way towards applications of quantum structures in optoelectronics and in bio-sensing and contributed to the development of nanotechnology. The major new challenge is in accessing the structural, electronic and optical properties of quantum dots on a nanoparticle scale in order to understand complex relationships between structural motifs and their contributions to the relevant physical (e. g. optical and electronic) properties. Here we demonstrate that a combination of molecular dynamics simulations and optically-detected x-ray absorption spectroscopy shows sufficient sensitivity to distinguish between regions contributing to the luminescence signal in oxygen and hydrogen terminated Ge quantum dots, thus potentially providing a sub-nanoparticle resolution. 
\end{abstract}

%\keywords{}

\maketitle

%\section{}
%\subsection{}
%\subsubsection{}
It is well-known that the reduction of materials size down to the nano-scale can have a significant impact on physical properties of the material. For semiconductors this can generally be understood in terms of the quantum confinement effect - a condition where the geometric size decisively affects a variety of physical parameters \cite{aliv1,aliv2,brus1}. The concept is elegant, but not easy to probe directly in many cases due to the difficulties in observing quantum dots (QDs) in an idealised state that can be readily compared with a corresponding theoretical or a computational model. A good example of this is porous nanocrystalline Si (\textit{p}Si), the unusual optical properties of which were discovered in 1990 by Canham \cite{ref1,ref2}. The intense visible photoluminescence (PL) observed in \textit{p}Si was originally attributed to the quantum confinement effect, but almost immediately another point of view was voiced \cite{ref3,ref4,ref5,ref6} suggesting the effect was due to the surface or silicon compounds of molecular nature. Two decades of intense research that followed could not provide unequivocal evidence and eventually led to the development of a model of Si QDs that includes core, surface, and interfacial regions \cite{ref7}. The recent contributions to the discussion includes rather complex PL measurements on single Si QDs in a silicon oxide environment \cite{ref6} and in magnetic fields \cite{nature2008}. Crucially, none of the approaches so far have been able to provide \textit{direct} evidence of a connection between optical signal and the underlying atomic structure and it seems that the  research into Ge QDs is suffering from a similar fate. Several preparation routes have been reported \cite{geprep1,geprep2,geprep3,geprep4} and PL was observed in the region between 400 nm and 1000 nm. The exact origins of the observed PL are yet to be established with some reports suggesting significant influence of surface effects \cite{geSolution}. 

Germanium is a close structural and electronic analogue to Si and there has been significant interest in understanding the optical properties of Ge QDs \cite{ref12,ref13,ref14}. The vast majority of early studies looked at embedded Ge QDs due to limited success in preparing free-standing samples with controlled surface termination. Only recently advances in colloidal synthesis \cite{collReview} provided access to free-standing Ge QDs. It has been established \cite{ref14,ref15} that in most cases PL observed in embedded Ge QDs can be attributed to oxide related species which is mostly related to the preparation techniques (e. g. implantation into an oxide \cite{ref16}, reduction from Ge oxides \cite{ref17}, etc.). Recently, a comprehensive XAS study has been conducted \cite{ref18} of nano-size effect on the structural properties of Ge QDs embedded in a silica matrix which demonstrated the formation of a disordered region between the nano-crystalline core and silica matrix. Again, the role of the matrix in the formation of these regions and the effect of he matrix on the optical properties of embedded Ge quantum dots is unclear. The main challenge is in establishing a \textit{direct} link between the atomic structure and optical emission. Our own effort over the last several years \cite{ref19} resulted in samples that do not show significant evidence of Ge-based oxides as characterised by Raman spectroscopy \cite{ref20} and EDX analysis while infrared FTIR/Raman data \cite{ref20} show the presence of Ge hydroxide/hydride species.

\begin{figure}
\includegraphics{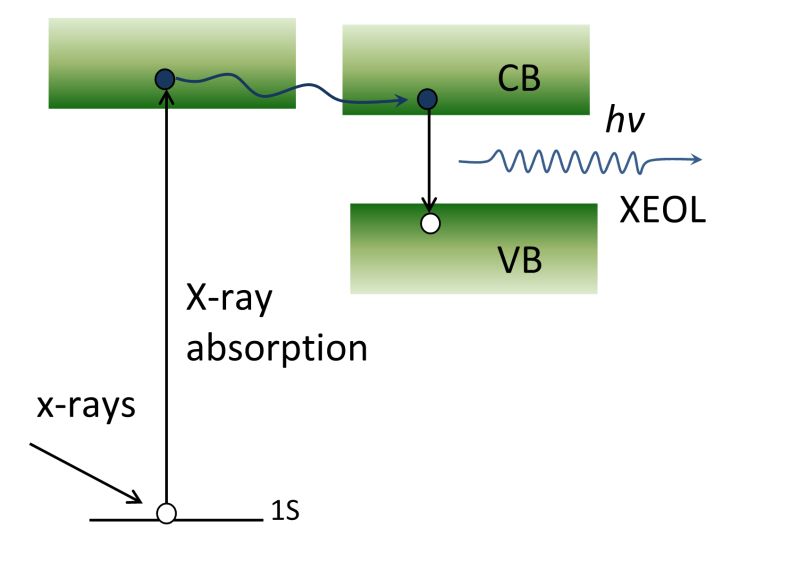}
\caption{\label{fig1} A schematic diagram of x-ray excitation-photoluminescence cycle in OD-XAS. An excitation from 1\textit{s} state to continuum followed by radiative recombination (XEOL) that carries information about XAS event.}
\end{figure}

Optically-detected x-ray absorption spectroscopy (OD-XAS) is a technique that enables structural data to be obtained directly from x-ray excited optical luminescence (XEOL) and has already been used to address the origins of PL in \textit{p}Si \cite{ref6,ref8,ref9,ref10,ref11}. OD-XAS is based on the XEOL emitted being sensitive to the photoelectrons that are generated in the x-ray absorption process near and above the x-ray absorption edge of  an element (see Fig. \ref{fig1}). The related x-ray absorption signal is measured by recording the integral photoluminescence yield within a selected wavelength range. Thus, it is sensitive to a subset of sites responsible for the light emission. Even so, OD-XAS has did not provide an unambiguous answer as to the origins of PL in \textit{p}Si. One of the difficulties is in the sensitivity of the OD-XAS method to the sample preparation \cite{ref12} which precluded wider use of the technique. Another difficulty in assessing spatial sensitivity of this method on the scale of a few nm \cite{ref8,ref9}. Here we turn our attention to the structural origin of visible PL observed in a series of Ge QDs:(i) hydrogen-terminated surface; (ii) oxygen-terminated surface; (iii) embedded into SiO$_{2}$ matrix. Samples in this study have been prepared by sol-gel synthesis \cite{ref_solgel} and by etching \cite{ref19,ref20}. We demonstrate that by combining the OD-XAS and molecular dynamics simulations it is possible to extract local structural information on a light emitting site and to obtain the details of the structural morphology. 

Free-standing Ge QDs were prepared by etching \cite{ref19,ref20} and silica-embedded Ge quantum dots were prepared by the sol-gel synthesis \cite{ref_solgel}. Surface of some of the samples prepared by etching was oxidized by exposure to air for two weeks  prior to x-ray absorption experiments. No further treatment has been done of SiO$_2$ encapsulated samples prepared by the sol-gel synthesis. The size of the Ge QDs was evaluated from TEM and from ambient Raman measurements using the relationship described previously \cite{ref25} and was found to be between 5 nm and 9 nm depending on the sample. 

OD-XAS experiments using x-ray excited optical luminescence (XEOL) signal have been conducted at beamline B18 at Diamond Light Source. All data were collected at Ge K-edge ($\approx$ 11 keV) with samples prepared to utilise the advantages of the thin limit \cite{ref_odxas_japs} to ensure consistensy between transmission and opticlly-detected signals. The detection system consisted of a Triax 190 spectrometer equipped with a Synapse CCD and a Newport VIS Femtowatt photoreceiver. A Hamamatsu R3809U-50 MCP photomultiplier was used for low signal conditions. The light was delivered to the spectrometer using an optical fiber. All experiments have been conducted at low temperature (\textit{T} = 100 K) using a cryojet system. OD-XAS data were reduced by PySpline \cite{ref23} and analysed using EXCURVE \cite{ref24} and FEFF \cite{ref_feff} codes. The value of "Goodness of fit" was used to identify the best fit to the data using radial distribution functions generated by molecular dynamics and is defined according to Lyttle et al.\cite{lytle}:
\begin{equation}
\varepsilon ^2 = \frac{1}{N_{ind} - p}\frac{N_{ind}}{N} \sum_{i=1}^{N} \left(\frac{\chi_i^{exp} - \chi_i^{theor}}{\sigma_i^{exp}}\right)^2
\end{equation}
where, $N$ is total number of points, $N_{ind}$ is the number of independent data points, $p$ number of parameters being refined, $\chi_i^{exp}$ and $\chi_i^{theor}$ are the experimental data and the theoretical model respectively, and $\sigma_i^{exp}$ is the the standard deviation for each data point, obtained by averaging several spectra. Transmission EXAFS data were used for reference. 

\begin{figure}
\includegraphics{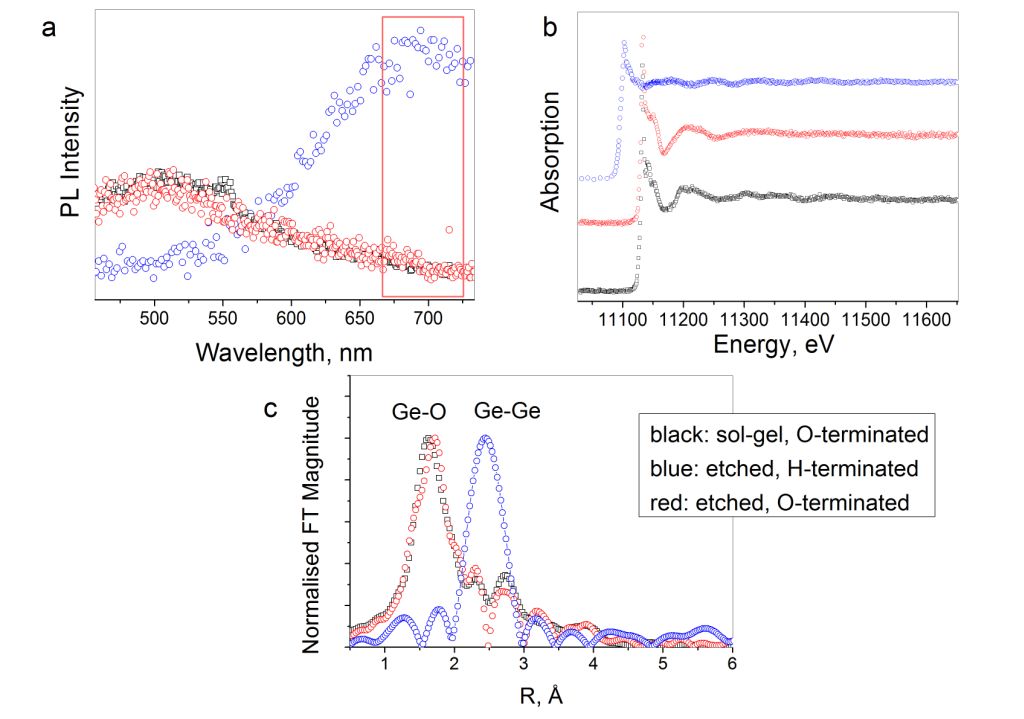}
\caption{\label{fig2} XEOL, corresponding OD-XAS and structural data for Ge QDs. \textbf{a}, XEOL data used to extract structural information. Red rectangle indicates the exact position of the band used to collect OD-XAS. \textbf{b}, OD-XAS data used to obtain structural information. Data for oxygen-terminated samples are shifted along x-axis for clarity. \textbf{c}, Structural data extracted from XEOL presented as normalized magnitude of Fourier transform obtained from OD-XAS data (see Methods section). Peaks correspond to nearest-neighbors around a Ge atom.}
\end{figure}

The results of analysis of OD-XAS data collected at the Ge K-edge ( $\approx$ 11 keV) for QDs with various surface termination are shown in Fig. \ref{fig2}. One can clearly observe the difference in the magnitude of the Fourier transform (FT) of the extended x-ray absorption fine structure (EXAFS) signal between samples. We can also conclude that in oxygen-terminated samples it is the oxygen-rich surface that gives a major contribution to the light emission as we see virtually no signal of the coordination shell corresponding to Ge atoms (a coordination shell at around $2.44$ \AA\ ). From this point on we only analysed the data for the hydrogen-terminated sample as data indicated that PL is associated with pure Ge rather than oxide species. The EXAFS analysis reveals a single peak at $R = 2.44\pm 0.01$ \AA\ from the central atom which is consistent with the corresponding value for the diamond-type structure of \textit{c}-Ge. However, we did not observe a multi-shell structure in the magnitude of FT that is common for \textit{c}-Ge at these temperatures \cite{ref26}. The lack of a structural signal beyond the first shell is usually an indication of a topologically disordered (e.g. amorphous) structure, but may also indicate the effect of surface disorder that can play an important role in small systems\cite{ref28,ref30,surface1}. For a single quantum dot the surface layer would most certainly be terminated with hydrogen atoms, but these are too weakly scattering to be observed in EXAFS. A reduced coordination number in the first shell can be an indication that structure is at the surface, but extracting accurate coordination from OD-XAS with sufficient accuracy can be a challenge due to complex nature of electron excitation-de-excitation processes \cite{ref_odxas_japs,ref_odxas_goulon,ref_odxas_pettifer,ref8}. At this point a source of extra information about regions contributing to the light emission is required in order to resolve this problem in analysis of OD-XAS data. In the following analysis we used a model of a single QD of an appropriate size (5 nm) to establish if we can further localise the source of the light emission. 

\begin{figure}
\includegraphics{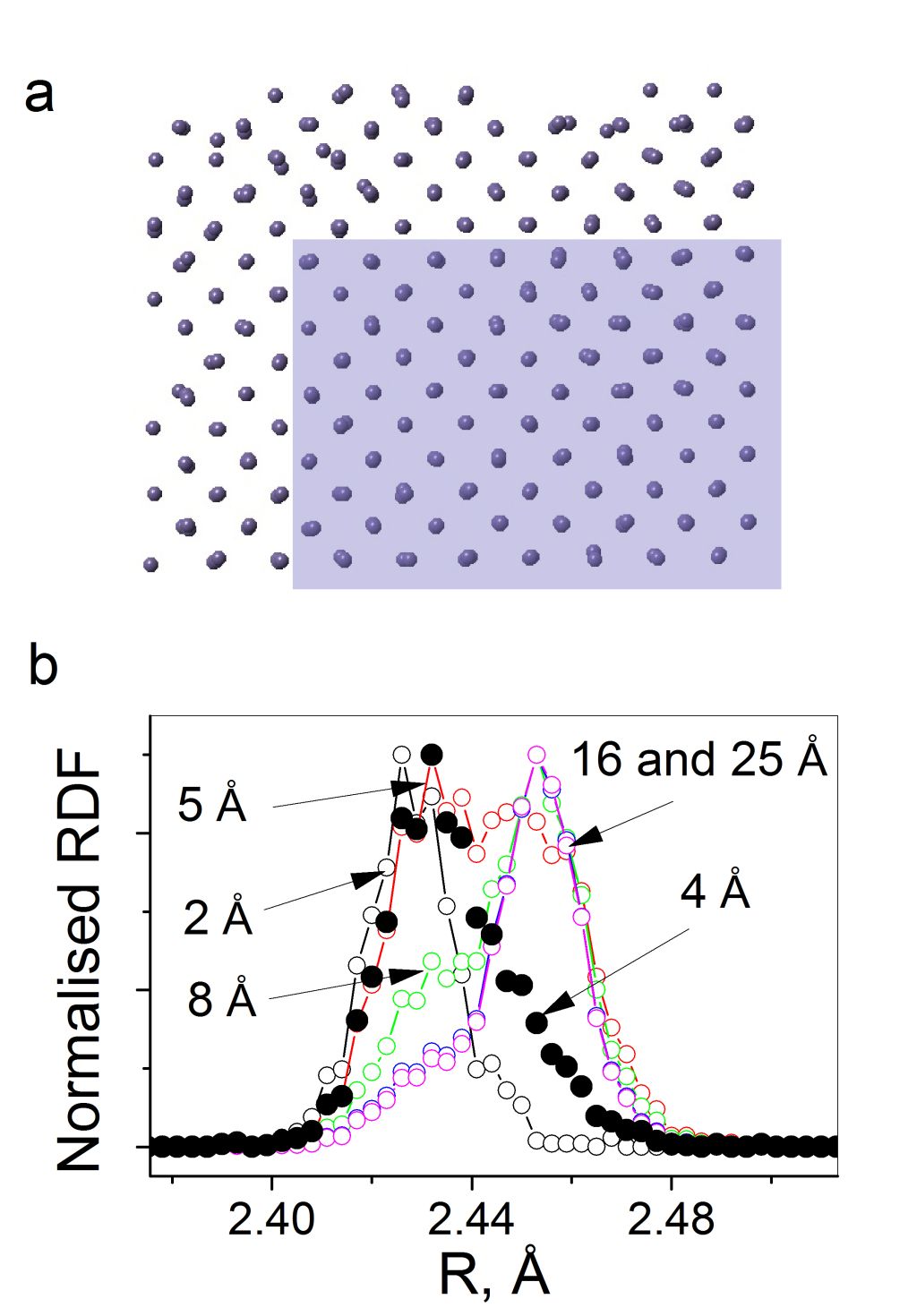}
\caption{\label{fig3} Results obtained from molecular dynamics simulations. \textbf{a}, A part of a 5 nm Ge particle generated by molecular dynamics simulations. Unshaded area corresponds to $\approx$ 5\AA\ surface layer and shows clear signs of disorder. \textbf{b}, RDFs extracted as a function of distance $d$ from surface. Numbers in \AA\ designate a corresponding RDF.}
\end{figure}

To understand the effect of reduced dimensionality on the OD-XAS signal, we have performed molecular dynamics (MD) simulations of bulk and nanoparticle Ge. We  used DL$\_$POLY MD package \cite{dlpoly}, and employed the highly successful and widely used environment-dependent Tersoff potential \cite{tersoff}, particularly well suited for surface simulations where the atomic environment is different from that in the bulk. The system size was approximately 50 \AA\ (as in experiments) involving about 6,000 atoms. Bulk configuration with periodic boundary conditions was equilibrated for 50 ps at 100 K. To simulate a finite-size Ge nanoparticle with a free surface the cell size was increased by 20 \AA\ - the distance larger than the potential cutoff. This resulted in atoms in the outmost layers to be connected to vacuum, and induced surface relaxation (see Figure \ref{fig3}). Similar surface relaxation has been reported before based on a combination of empirical and first-principles molecular-dynamics techniques \cite{ref28,ref30,surface1}. To quantify the effect of surface relaxation, we have extracted the radial distribution function (RDF) for both bulk and free surface structures at both temperatures. We extracted RDFs as a function of distance $d$ from the surface towards the center of the particle. In Figure \ref{fig3}, we plot RDFs for different values of $d$, and observe a clear bi-modal distribution of distances as $d$ increases towards the centre of a nanoparticle. This bi-modal distribution is a non-trivial result and suggests that a surface/interface layer is structurally different from the crystalline core. This in turn would suggest a distinctive difference in the electronic and optical properties of the surface/interface layer as compared to that of the core. 

It is not possible to observe a level of structural detail obtained from MD  in OD-XAS since MD data give positions of nuclei while EXAFS is sensitive to the electron density distribution. As a consequence, the associated mean-square relative displacements obtained from OD-XAS are too large and the details are masked due to thermal contribution. However, it should still be possible to observe shortening of the average interatomic distance in the data extracted from OD-XAS if only the surface states are responsible for the light emission. We observe no obvious shortening of distance in our OD-XAS experimental data. However, the cumulant analysis of the first neighbour Ge-Ge peak ($R = 2.44\pm 0.01$ \AA ) in OD-XAS data indicates a non-zero value of the third cumulant ( $0.005\pm 0.003$ \AA $^3$, skewness of the peak) while the third cumulant in bulk crystalline reference sample is close to zero ($0.0003\pm 0.0018$ \AA $^3$). This peak skewness indicates that there may be a contribution from more than one shell of atoms. 

\begin{figure}
\includegraphics{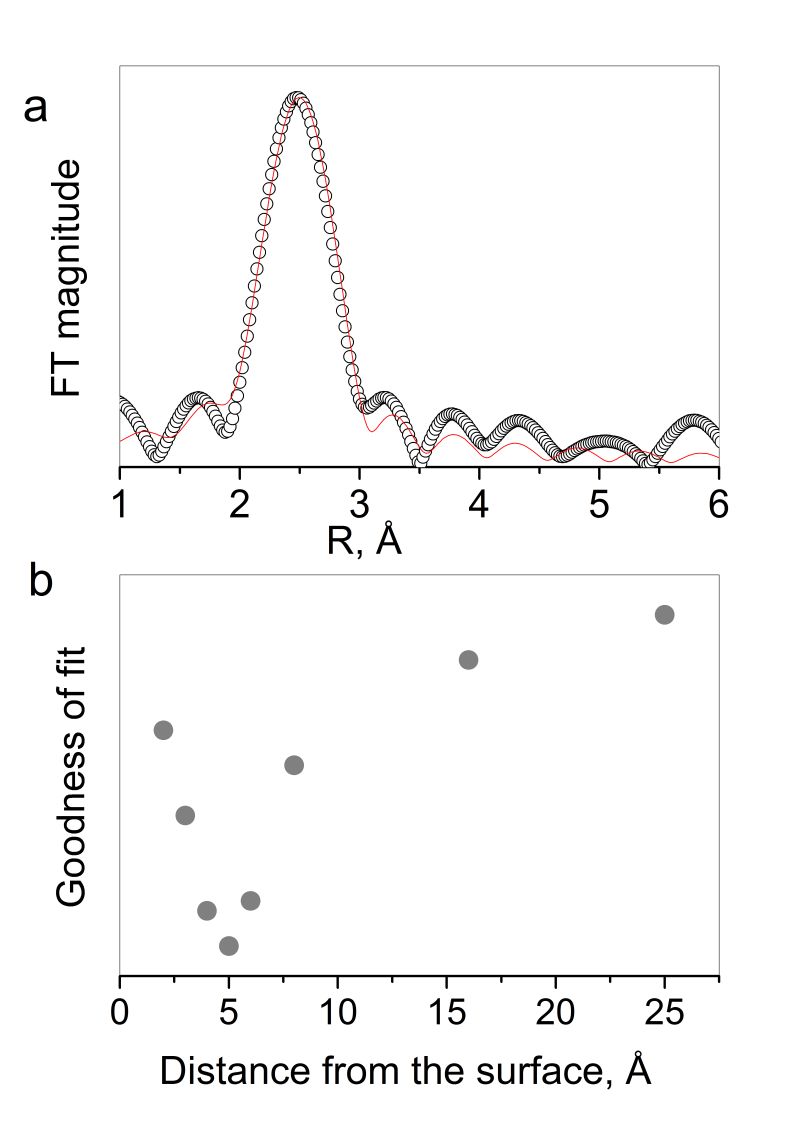}
\caption{\label{fig4} Results of the fitting of molecular dynamics based models to OD-XAS derived data. \textbf{a}, Experimental data and the best fit. \textbf{b}, Goodness of fit for various RDFs calculated as a function of distance from the surface towards the center of the QD.}
\end{figure}

The difficulty in extracting much more detailed structural information from a single peak obtained from OD-XAS is that the number of relevant structural parameters (e.g. interatomic distances, numbers of neighbours, Debye-Waller factors) is limited by the well-known Nyquist theorem \cite{Nyq}. Parameters can also be highly correlated, leading to over-determination of the fit and increased errors. Therefore, in order to investigate the origins of peak skewness and to recover the sub-structure responsible for the observed OD-XAS signal we used RDFs obtained from MD simulations.  This allowed us to reduce the number of variable parameters to just two during refinement: (i) a single nearest neighbour number and (ii) a single Debye-Waller factor. The ratio between surface and bulk contributions to the RDF was fixed based on MD results. The information obtained from MD (interatomic distances and ratio of nearest neighbours) were then used to generate an EXAFS signal and to compare the model with the experiment for a number of RDFs between 2 \AA\ and 25 \AA\ . The result of this comparison can be seen in Figure \ref{fig4}. We found a clear minimum in the value of the fit index (characterizes the "goodness" of a fit) for the RDF corresponding to the layer of 5 \AA\ within the model of a single Ge QD. This indicates that the origin of light emission can be localized to a substructure up to 5 \AA\ from the surface towards the center of a particle within our model.  The results suggest that it's possible to identify structural motifs responsible for the light emission in a nanoscale system. We have to note that the level of structural localisation we report here is based on a specific morphological model - that of a layered QD. However, the result has a general validity beyond the model in that it is disordered structural component that contributes to the light emission. The point is that MD simulations allow one to obtain a detailed structural model that can be further used to examine structural and optical properties of the system. 

In summary, we demonstrate that a combination of molecular dynamics simulations and OD-XAS shows sufficient sensitivity to identify the structural region contributing the light emission in Ge QDs, thus potentially providing sub-nanopaticle resolution.  We show that in Ge QDs the structural contribution to photoluminescence crucially depends on the surface termination. In samples where the surface is oxidised there is a clear contribution from the oxide to the light emission. In hydrogen-terminated samples with the aid of molecular dynamics simulation we show that the disordered region possibly located at the interface between the core and the surface plays a key role in the light emission. These findings suggest that a quick and assessment can be made of the relationship between preparation conditions, related structure, and relevant optical properties in light emitting quantum dots.   
% If in two-column mode, this environment will change to single-column
% format so that long equations can be displayed. Use
%\begin{widetext}
%\end{widetext}

% figures should be put into the text as floats.
% Use the graphics or graphicx packages (distributed with LaTeX2e)
% and the \includegraphics macro defined in those packages.
% See the LaTeX Graphics Companion by Michel Goosens, Sebastian Rahtz,
% entire page. There is no need to do explicit centering.

% Surround figure environment with turnpage environment for landscape
% figure
% \begin{figure}
% \includegraphics{fig1.eps}%
% \end{figure}

% tables should appear as floats within the text
% The ruledtabular enviroment adds doubled rules to table and sets a
% reasonable default table settings.
% Use the table* environment to get a full-width table in two-column
% Add \usepackage{longtable} and the longtable (or longtable*}
% environment for nicely formatted long tables. Or use the the [H]
% placement option to break a long table (with less control than 
% \begin{table}%[H] add [H] placement to break table across pages
% \begin{ruledtabular}
% Lines of table here ending with \\
% \end{tabular}
% \end{ruledtabular}
% \end{table}

% Surround table environment with turnpage environment for landscape
% table
% \begin{table}
% \begin{ruledtabular}
% \end{tabular}
% \end{ruledtabular}
% \end{table}

%\appendix
%\section{}

\begin{acknowledgments}
The beamtime provision by Diamond Light Source are gratefully acknowledged. W. Little is grateful to South East Physics Network for financial support.
\end{acknowledgments}

%\bibliography{odxas_refs2012}

\end{document}